  \documentstyle[prl,aps,epsf,two column]{revtex}

  \include{psfig}
  \begin{document}
  \title{Giant lasing effect in magnetic nanoconductors}
  \author{A. Kadigrobov$^{1,2}$,
  \fbox{Z. Ivanov} and T. Claeson$^{1}$}
  \address{
  $^{1}$Department of Microelectronics and Nanoscience, Chalmers
  University of Technology,
  SE-412 96 G\"oteborg, Sweden\\
  }
  \author{R. I. Shekhter$^{2}$, and M.~Jonson$^{2}$
  }
  \address{
  $^{2}$Department of Applied Physics,
  Chalmers University of Technology and G\"oteborg University,
  SE-412 96 G\"oteborg, Sweden
  }


  \date{\today}
  \maketitle
  \begin{abstract}
  We propose a new principle for a compact solid-state laser in the 1-100
  THz regime.
  This is a frequency range where attempts to
  fabricate small size lasers up till now have met severe technical
  problems.
  The proposed laser is based on a new mechanism for creating spin-flip
  processes
  in ferromagnetic conductors. The mechanism is due to the interaction of
  light
  with conduction electrons; the interaction strength, being proportional
  to
  the large exchange energy, exceeds the Zeeman interaction by orders of
  magnitude.
  On the basis of this interaction, a giant lasing effect is predicted
  in a system where a population inversion has been created by tunneling
  injection of
  spin-polarized electrons from one ferromagnetic conductor to another ---
  the
  magnetization of the two ferromagnets having different orientations.
  Using experimental data for ferromagnetic manganese perovskites with
  nearly
  100\% spin polarization we show the laser frequency to be in the range
  1-100 THz.
  The optical gain is estimated to be
  of order $10^7$ cm~$^{-1}$, which exceeds
  the gain of conventional semiconductor lasers by 3 or 4 orders of
  magnitude.
  A relevant experimental study is proposed and discussed. \\
  \end{abstract}
  
\begin{figure}
  \centerline{\psfig{figure=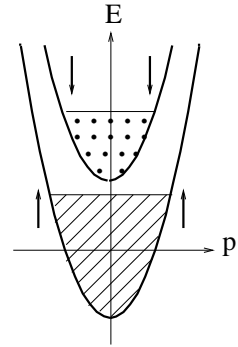,width=3cm}}
  \vspace*{2mm}
  \caption{Schematic representation of the band structure of a
  magnetically
  ordered conductor. The hatched region corresponds to an equilibrium
  distribution of spin-up electrons in the lower band, while the dotted
  region
  indicates a non-equilibrium distribution of "hot" spin-down electrons
  in the upper band. Arrows show electron spin directions in the bands}
  \label{hot}
  \end{figure}

\begin{figure}
  \centerline{\psfig{figure=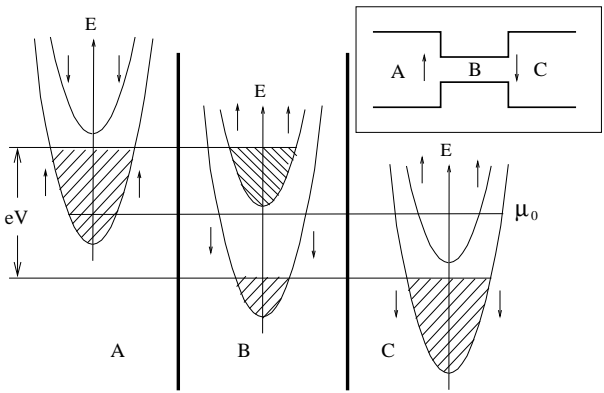,width=8cm}}
  \vspace*{2mm}
  \caption{Schematic illustration of a population inversion in a magnetic
  conductor
  (region B).
  A bias voltage $V$ is applied between magnetic conductors
  ($A$ and $C$)
  which have opposite directions of their magnetizations (the thick 
  vertical lines represent
  potential barriers). Arrows show the
  electron spin
  directions in the electron energy bands; $\mu_{0}$ is the equilibrium
  chemical potential
  (in the absence of bias). Population inversion requires the bias voltage
  to be $V > (2 I - \mu_0)/e$. The insertion shows a possible realization 
  of a structure
  in which two magnetic conductors with opposite magnetization directions
  are coupled through a microbridge (B).}
  \label{population}
  \end{figure}

  The ability to synthesize magnetically ordered, layered
  conductors with nearly 100$\%$ spin-polarization
  of the conducting electrons \cite{Tokura}, has opened up a new field 
  in solid state
  physics, the field of "spintronics" \cite{Prinz,Wolf}.
  Spin-dependent tunneling of electrons is one phenomenon that has already
  found commercial applications based on the resulting "giant"
  magnetoresistance of certain layered structures \cite{Prinz,Wolf}.
  Other applications are bound to follow.
  The possibility to control, by means of a bias voltage, not only the
  energy but also the spin of electrons injected into a magnetic conductor
  makes it feasible to investigate the properties of highly excited
  spin-polarized electrons. An example of such a system is presented in
  Fig.\ref{hot}, where the hatched region corresponds to an equilibrium
  distribution of (spin-up) electrons in a spin-polarized conductor.
  The dotted area marks a non-equilibrium distribution of "hot"
  (spin-down) electrons.
  Relaxation of the spin-down electrons to an equilibrium configuration
  requires
  spin-flip processes and is therefore completely blocked if such
  processes
  are not allowed. In the presence of such a "spin lock" against
  relaxation,
  highly excited states in the material may have a long lifetime, which
  may in turn
  determine novel "spintronics" effects in spin-polarized conductors.
  The objective of this Letter is to demonstrate how electromagnetic
  radiation
  may remove the spin-lock effect and to demonstrate some important
  consequences
  of this effect for spintronics. We will show that the radiation makes
  the blockade
  of relaxation weaker through
  its coupling to the exchange interaction in magnetically ordered
  conductors. This comes about via the dependence of the exchange constant
  on
  the momenta of the conduction electrons.
  As a result a lasing effect is shown to occur in systems where an
  inverted
  electron population has been created by the tunneling injection of
  spin-polarized
  electrons from one ferromagnetic conductor to another (the orientation
  of the
  magnetization being different in the two ferromagnets).
  An example of such a system is presented in Fig.(\ref{population}).
  Our estimations show that a laser with an optical gain that
  exceeds the gain of conventional semiconductor lasers by three or
  four orders of magnitude can be built and argue that laser action can be
  achieved provided care is taken to design the system so that the lasing
  region is not too much heated.
  The frequency of these lasers can be in a wide range that
  includes the interval 1 - 100 THz.
  %

  The Hamiltonian for the electrons
  in a magnetically ordered conductor can be written as
  \begin{equation}
  \hat{H}_0 = \hat{\sigma_0} \frac{\hat{p}^2}{2m*}
  -\hat{\mbox{\boldmath$ \sigma$}}{\bf I}\,,
  \label{ham0}
  \end{equation}
  where $m*$ is the effective mass, $ \hat{{\bf p}}$ is the momentum
  operator,
  $\hat{\mbox{\boldmath $\sigma$}}$ are Pauli matrices, $\sigma_0$ is
  the $2\times 2$ unity matrix, and $I$ is the exchange energy.
  According to Eq.(\ref{ham0}) the dispersion law for electrons with
  spins pointing up/down is
  \begin{equation}
  E_{\uparrow \downarrow}\left(p \right) =\frac{p^2}{2m*} \mp I
  \label{dispersion}
  \end{equation}

  We will deal with a system, schematically presented in
  Fig.\ref{population}, in which two potential barriers divide the 
  magnetic conductor into three parts;
  the magnetization of two adjacent magnetic conductors (regions A and B)
  are
  pointing in opposite directions.
  A bias voltage $V$ is applied between regions A and C. We assume that
  the spin relaxation time $\nu_s^{-1}$ \cite{absorpion}, the time of 
  energy relaxation without
  changing the electron spin direction $t_E$ and the time of electron 
   tunneling $t_{tun}$
  obey the inequality $t_E \ll t_{tun} \ll \nu_s^{-1}$.
  In the absence of spin-flip processes, the energy relaxation of injected 
  spin polarized electrons
  creates a non-equilibrium state in which equilibrium is established 
  only within each group of electrons with a fixed spin polarization.
  Therefore, in region B electrons in the spin-up and spin-down 
  energy bands are in local equilibria corresponding to the different
  chemical potentials $\mu_\uparrow $ and $\mu_\downarrow$, 
  respectively, while the system as a whole is far from equilibrium.

  According to Eq.(\ref{dispersion}) the energy conservation law for
  vertical transition
  of electrons with emission of photons of frequency $\omega$
  does not depend on the electron momentum:
  \begin{equation}
  E_{\uparrow }\left(p \right) - E_{\downarrow}\left(p \right) - \hbar
  \omega = 2 I - \hbar \omega
  \label{conserv}
  \end{equation}
  It follows that for $\omega = 2I/\hbar$ all "hot" electrons are in
  resonance
  with the electromagnetic field, and hence stimulated emission of light
  due to transitions of
  electrons from filled states of the upper band to the empty states of
  the lower band is
  possible for all electrons in the energy range
  $\mu_\uparrow - \mu_\downarrow $.
  As seen in Fig.\ref{population}, under the neutrality condition
  $N_\uparrow +N_\downarrow =N^{(0)}$ ($N_\uparrow$ and $N_\downarrow$ 
  are the densities of spin-up and spin-down electrons,
  $N^{(0)}$ is the equilibrium electron density)
  the population inversion
  needed for lasing requires a large enough bias voltage,
  $V > (2 I - \mu_0)/e$.

  The conventional Zeeman term $\hat{H}_Z= g\mu_B {\bf
  H}\hat{\mbox{\boldmath$ \sigma$}}$
  describing interaction between the (hot) electrons and an
  electromagnetic field does
  provide a mechanism for stimulated radiative transitions between the
  energy bands
  containing electrons with opposite spin
  directions. However, it is relatively small in magnitude and it is not
  the most important
  mechanism.
  For ferromagnets, we would like to suggest a much more effective
  mechanism of
  interaction between light and conduction electron spins.
  This mechanism is based on the dependence of the exchange energy
  (\ref{ham0}) on
  the momentum ${\bf p}$ of the conduction electron. The momentum 
  dependence has to
  do with the
  overlap of the wavefunctions of the conduction electron and the magnetic
  sub-system
  (see, e.g., \cite{Zeiger}). It is determined by the value of
  $ p a/\hbar$, where $a$ is the characteristic size of the orbital (that
  is $a$ is an
  atomic-scale length). This is why it varies with the momentum of the
  conduction electron.
  In the absence of an electromagnetic field the Hamiltonian which
  describes this situation
  can be written as
  \begin{equation}
  \epsilon(\hat{{\bf p}}) = \hat{\sigma_0} \frac{\hat{p}^2}{2m*} -
  \hat{\mbox{\boldmath$ \sigma$}}{\bf I}(\hat{{\bf p}}) \,.
  \label{implicit}
  \end{equation}
  In the presence of an electromagnetic field with vector potential ${\bf
  A}
  $ the momentum operator $\hat{{\bf p}}$ in Eq.(\ref{implicit}) must be
  changed
  to $\hat{{\bf p}} -(e/c) {\bf A}$. Expanding in powers of $(e/c) {\bf
  A}$ one gets
  an effective Hamiltonian $\hat{H}_{eff} =\epsilon(\hat{{\bf p}}) +
  \hat{H}_{eff}^{(1)}$
  where the perturbation Hamiltonian \cite{spin-orbit} is
  \begin{equation}
  \hat{H}_{eff}^{(1)} =-\frac{e}{2c}\hat{\mbox{\boldmath$ \sigma$}} (A_i
  \frac{\partial{\bf I}}{\partial p_i} + \frac{\partial{\bf I}}{\partial
  p_i} A_i)_{{\bf p} =\hat{{\bf p}}}
  \label{effham}
  \end{equation}
  In Eq.(\ref{effham}) we have omitted terms that do not flip spins;
  summation over double indices is implied: $a_ib_i \equiv {\bf a}{\bf
  b}$.
 \begin{figure}
  \centerline{\psfig{figure=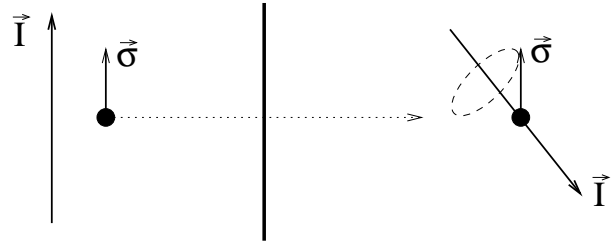,width=8cm}}
  \vspace*{2mm}
  \caption{Schematic illustration of how electrons are injected into the
  active region
  for the case that the adjacent ferromagnets have different magnetization
  directions (shown with long arrows; electron spins, $\vec{\sigma}$, are
  shown
  with short arrows with black dots at the end).
  An electron from the left-side ferromagnet with its spin parallel to
  the
  magnetization direction passes through a sharp boundary (shown as a
  vertical
  thick line) between the ferromagnets without changing spin direction.
  In the right-side
  ferromagnet it emerges with its spin
  in a superposition of the spin-up and spin-down states. The classical
  precession of the spin is indicated by a dotted ellipse.}
  \label{injection}
  \end{figure}


  If the injected electrons are prepared in such a way that their spins
  are not
  parallel to the magnetization in the active region B (see Fig.\ref{population}), 
  the Hamiltonian
  (\ref{effham})
  produces spin-flips and hence stimulates the needed radiative
  transitions of hot
  electrons in the upper band to the lower energy band. This process is
  illustrated in Fig.\ref{injection}, where an electron
  (with its spin parallel to the magnetization) is impinging on the
  boundary
  from the left, passes through the boundary and is scattered into a
  quantum superposition of spin-up and spin-down states in the active
  region to the right of the boundary. The wave
  function of the electron in the active region B is
  \begin{equation}
  \Psi ({\bf r}) =
  e^{i{\bf p}_\perp{\bf r}_\perp}\sum_{k=1}^2 a_k \left( \begin{array}{c}
  1 \\
  I_+/(I_z +(-1)^k I) \end{array}\right)e^{i p_k x}
  \label{wave}
  \end{equation}
  where the $x$-axis is perpendicular and the $y$- and $z$-axes are parallel to the 
  boundary; the projections of the electron momentum and the coordinate on the
  boundary are ${\bf p}_\perp = (0, p_y,p_z)$ and ${\bf r}_\perp
  =(0,y,z)$, while
  $p_{1,2} = \sqrt{2m*(E\mp I) -p_\perp^2 }$ and $I_+ = I_x + i I_y$;
  coefficients
  $a_{1,2}(E)$ are found by matching the wave functions of the electron in
  the active region (the right-hand side of Fig.~\ref{injection}) and in
  the
  injecting region (left-hand side of Fig.~\ref{injection}) at the
  boundary
  (we do not present their explicit form here).

  Using Eq.(\ref{wave}) as the initial proper state $\Psi_i ({\bf r})$
  belonging to the initial energy $E_i$ and the final state $\Psi_f ({\bf
  r})$
  belonging to the final energy $E_f$ one sees that matrix element
  $\left<\Psi_f ({\bf r})\right|\hat{H}_{eff}^{(1)}\left|\Psi_i ({\bf
  r})\right>$
  (the probability amplitude for a radiative electronic transition between
  the unperturbed
  energy bands) is not zero if ${\ p}_1(E_i)={\ p}_2(E_f)$. From here it
  follows
  that the difference between the initial and the final energies should be
  $E_i({\bf p}) -E_f({\bf p})= 2I({\bf p})$.

  Taking such wave-functions as the initial and the final states and
  performing standard calculations \cite{Agrawal} using
  Eq.(\ref{effham}) and Fermi's golden rule one gets the stimulated
  transition rate per unit volume of active material as
  \begin{eqnarray}
  R_{st} = \frac{32\pi e^2 \mu}{ n^2}\frac{|{\bf I}^\prime|^2}{\omega
  }\frac{ \hbar \nu_s \left( b_1 N_\uparrow - b_2 N_\downarrow\right)
  }{\left(\hbar \omega -2I \right)^2 + \left(\hbar \nu_s\right)^2}
  \label{ratesf1}N_{p}\\
  \times \left\{
  \left(
  \left[\mbox{\boldmath$ \epsilon$}_a [\mbox{\boldmath$ \epsilon$}_{dr}
  \cdot \mbox{\boldmath$ \epsilon$}_a]\right]\cdot \mbox{\boldmath$
  \epsilon$}_{i}
  \right)^2 +
  \left(
  \left[\mbox{\boldmath$ \epsilon$}_{dr} \cdot \mbox{\boldmath$
  \epsilon$}_a\right]\cdot \mbox{\boldmath$ \epsilon$}_{i}
  \right)^2
  \right\} \nonumber
  \end{eqnarray}
  Here $\mu$ and $n$ are the magnetic permeability and the refractive
  index of
  the medium, respectively, $N_p$ is the photon density, $\omega$ is the
  photon
  frequency, the constants $b_1 $ and $b_2$ are of order unity,
  the unit vectors $\mbox{\boldmath$ \epsilon$}_a$
  and
  $\mbox{\boldmath$ \epsilon$}_{i}$ are directed along the
  magnetizations
  in the active (right-hand side of Fig.\ref{injection}) and injection
  (left-hand side of Fig.\ref{injection}) regions, respectively, while
  $\mbox{\boldmath$ \epsilon$}_{dr}$ is parallel to the vector
  ${\bf I}^\prime \equiv e_i \partial {\bf I}/\partial p_i$, where
  ${\bf e}$ is the unit polarization vector in the direction of
  the vector potential ${\bf A}$.

  Eq.(\ref{ratesf1}) was derived under the assumption that
  $|I^\prime| p_{F\uparrow} < \hbar \nu_s$ ($p_{F\uparrow}$ is the Fermi
  momentum of electrons in the upper band), that is the additional
  dispersion
  caused by the dependence of $I$ on the electron momentum ${\bf }$ (see
  Eq.(\ref{implicit}))
  is smaller than the broadening of the electron energy due to spin-flip
  processes.
  In the opposite limiting case $|I^\prime| p_{F\uparrow} \gg \hbar
  \nu_s$
  one has
  \begin{eqnarray}
  R_{st}= \frac{32 \pi}{\hbar \omega}\frac{e^2 \mu}{n^2} |{\bf
  I}^{'}|\left( b_3 N_\uparrow^{2/3} - b_4 N_\downarrow^{2/3}\right)
  \label{ratesf2}N_{p}\\
  \times \left\{
  \left(
  \left[\mbox{\boldmath$ \epsilon$}_a [\mbox{\boldmath$ \epsilon$}_{dr}
  \cdot \mbox{\boldmath$ \epsilon$}_a]\right]\cdot \mbox{\boldmath$
  \epsilon$}_{i}
  \right)^2 +
  \left(
  \left[\mbox{\boldmath$ \epsilon$}_{dr} \cdot \mbox{\boldmath$
  \epsilon$}_a\right]\cdot \mbox{\boldmath$ \epsilon$}_{i}
  \right)^2
  \right\} \nonumber
  \end{eqnarray}
  where the constants $b_{3,4} \sim 1$.

  One of the necessary conditions for the lasing effect to be realized is
  (see, e.g., \cite{Agrawal}):
  \begin{eqnarray}
  R_{st} =\nu_p N_p \,,
  \label{lasing}
  \end{eqnarray}
  where $\nu_{p}$ is the photon relaxation rate.
  We consider the case that damping of
  electromagnetic waves is mainly due to
  absorption by free charge carriers, the frequency of the photon
  relaxation
  being $\nu_{p}= 2 k \omega /n$ (see \cite{Blatt}).
  For estimating the parameters of the problem we use standard formulae
  for the refractive index $n$ and the absorption coefficient $k$ for
  metals
  subject to electromagnetic fields (see, e.g. \cite{Ziman}).
  One finds that
  \begin{eqnarray}
  \nu_{p}= \frac{\mu}{n^2} \frac{4 \pi \sigma (0)}{ 1+(\omega t_0)^2} \,.
  \label{2nk}
  \end{eqnarray}
  where $\sigma_0 $ is the static conductivity of the conductor, $t_0$ is
  the transport electron relaxation time.

  Using Eq.(\ref{ratesf1}) and Eq.(\ref{2nk}) one can
  rewrite
  Eq.(\ref{lasing}) as
  \begin{eqnarray}
  \frac{\left(N_\uparrow - N_\downarrow\right)}{N_\uparrow+N_\downarrow}
  \approx\frac{\hbar \nu_s}{2 I}\frac{p_0^2/m*}{\hbar t_0^{-1} }
  \frac{1}{1+(\omega t_0)^2} \,,
  \label{lasing1}
  \end{eqnarray}
  where $p_0 =\hbar/a \sim 10^{-19}$ erg sec/cm.

  It seems that for achieving the lasing effect the most favorable
  materials are
  ferromagnetic manganese perovskites with nearly 100\% spin polarization
  of
  the conduction electrons (\cite{Tokura,Park,Tokura2}). The high degree
  of polarization
  of the carriers permits the creation of a population inversion of the
  energy bands
  in the active region B (see Fig.\ref{population}). 
  Here and below we use
  experimental
  values of the needed parameters:
  the mean free path $l_0 = 1.4 \times 10^{-7}$ cm, the Fermi velocity
  $v_F \approx 10^{8}$ cm/s, $t_0 \approx 10^{-15} $ s and $m* = 0.3 m_e$
  where
  $m_e$ is the free electron mass, the number of carriers $\approx 3.4
  \times 10^{21}$,
  the resistivity $\rho \sim 10^{-4}\div 10^{-3}$ (\cite{comment}).
  Inserting these values into Eq.(\ref{lasing1}) one finds
  the lasing condition to be
  \begin{eqnarray}
  \frac{N_\uparrow -N_\downarrow}{N_\uparrow+N_\downarrow} \approx 5
  \frac{\hbar \nu_s}{I}
  \nonumber
  \end{eqnarray}
  For the case $|I^\prime| p_{F\uparrow} \gg \hbar \nu_s$ 
  (see Eq.(\ref{ratesf2})), the lasing
  condition
  Eq.(\ref{lasing}) is
  \begin{eqnarray}
  \frac{N_\uparrow^{2/3} -
  N_\downarrow^{2/3}}{N_\uparrow+N_\downarrow}\approx 0.5 \times 10^{-7}
  \,{\rm cm}
  \nonumber
  \end{eqnarray}

  From these equations it follows that the lasing condition $R_{st}=\nu_s
  N_p$ is easily
  satisfied since one needs
  $\hbar \nu_s/I $ to be less than $10^{-1}$ while the theoretical
  estimate of the spin
  relaxation rate $\nu_s$ gives the value $10^{-2}$ for this ratio.
  Estimations based on the above experimental values of the
  parameters
  show the optical gain to be $g_{opt} =(n/c)R_{st} \sim 10^7$
  cm$^{-1}$
  and the threshold current density to be
  $j_{th} = e l \nu_s N_\uparrow \sim 10^7 \div10^8 $ A/cm~$^{-2}$ if
  the length of
  the active region is $l = 10^{-5}$ cm. Estimations for $N_e \sim 10^{18}$
  cm~$^{-3}$ shows
  the optical gain and the threshold current to be $g_{opt} \sim 10^3
  \div 10^4
  $ cm~$^{-1}$
  and $j_{th} \sim 10^5$ A/cm~$^{-2}$.

  We predict an extremely large optical gain in systems with a high
  density of charge carriers. The price to be paid
  for the gain exceeding what can be achieved in semiconductors by 3 or
  4 orders
  of magnitude is the high currents needed for an effective tunneling
  pumping of
  the system. The current value $j = 10^6 \div 10^8 $ A/cm$^2$ seems to be
  very large for homogeneous bulk metals because of the accompanying Joule
  heating.
  Special measures are needed to avoid heating the active, lasing region.
  One solution to that problem is to arrange for the current injection to
  be
  inhomogeneous in space. This can be achieved
  if the magnetic conductors are electrically connected through a
  point contact. The spreading out of the current
  far from
  the narrow point contact provides for an efficient dissipation of heat
  \cite{Yanson}.
  A current density $j \sim 10^8$ A/cm$^2$ can be reached without
  significant heating of the
  contact region. On the other hand,
  the extremely large optical gain $g_{opt} \sim 10^7$ cm$^{-1}$ means
  that
  it is enough to have a small volume of active lasing region.
  Such a structure can be prepared on the basis of the technique suggested
  in Ref.~\cite{Zdravko} for fabrication of biepitaxial films of 
  $La_{0.7}Sr_{0.3}MnO_3$
  with
  $45^\circ$
  in-plane rotated domains.


  In summary we have proposed a new principle for a compact solid-state
  laser working
  in the 1-100 THz regime.
  The proposed laser is based on a new mechanism for creating spin-flip
  processes
  in ferromagnetic conductors. The mechanism is due to the interaction of
  light
  with conduction electrons; the interaction strength, being proportional
  to
  the large exchange energy, exceeds the Zeeman interaction by orders of
  magnitude.
  On the basis of this interaction, a giant lasing effect was predicted
  for systems where a population inversion can be created by tunneling
  injection of
  spin-polarized electrons from one ferromagnetic conductor to another ---
  the
  magnetization of the two ferromagnets having different orientations.
  Using experimental data for ferromagnetic manganese perovskites with
  nearly
  100\% spin polarization we show the laser frequency to be in the range
  1-100 THz.
  The optical gain is estimated to be
  of order $10^7$ cm~$^{-1}$, which exceeds
  the gain of conventional semiconductor lasers by 3 or 4 orders of
  magnitude.
  An experimental study based on a point contact geometry to avoid heating
  by the necessarily large injection currents was proposed and discussed.

  Acknowledgements.
  We thank L.Y. Gorelik, V. Kozub, G.D. Mahan and R. Gunnarsson for 
  useful discussions.
  Financial support from the Royal Swedish Academy of
  Sciences
  (AK) is gratefully acknowledged.



\begin{thebibliography}{99}
\bibitem{absorpion} The main mechanism for relaxation of hot electrons is 
  the spin-flip process caused by emission of spin-waves.
\bibitem{comment}  
  The values are taken for a compound similiar to the 
  perovskites manganites found to be nearly 100\% spin-polarized, see 
  \cite{Tokura2}.
\end{thebibliography}
  \end{document}